\documentclass[12pt,a4paper]{article}

\usepackage{latexsym,amssymb}
\usepackage{graphicx,color}
\usepackage{amsmath}
\usepackage{yhmath}
\usepackage{empheq}
\newcommand\sect[1]{\section{#1}\setcounter{equation}0}
\newcommand\ds{\displaystyle}
\newcommand\no{\nonumber\\{}}

\newcommand\eqnb{\begin{eqnarray}}
\newcommand\eqne{\end{eqnarray}}

\newcommand{\mf}{\mathfrak}

\newcommand{\C}{\ensuremath{\mathbb{C}}}
\newcommand{\R}{\ensuremath{\mathbb{R}}}
\newcommand{\Z}{\ensuremath{\mathbb{Z}}}

\usepackage{latexsym}

\newcommand\void[1]{}   

\newcommand{\al}{\alpha}
\newcommand{\be}{\beta}

\newcommand{\Tr}{{\mathrm{Tr}}}
\newcommand{\ii}{{\mathrm{i}}}

\newtheorem{theorem}{Theorem}

\newtheorem{lemma}{Lemma}

\title{On the unitarity of gauged non-compact\\ world-sheet supersymmetric WZNW models}
\author{Jonas Bj\"ornsson and Stephen Hwang\\
Karlstad University}

\begin{document}

{\center{{\huge On the unitarity of gauged non-compact\\ world-sheet supersymmetric WZNW models}\vspace{15mm}\\
Jonas Bj\" ornsson\footnote{jonas.bjornsson@kau.se} and Stephen 
Hwang\footnote{stephen.hwang@kau.se}\\Department of Physics\\Karlstad University
\\SE-651 88 Karlstad, Sweden}\vspace{15mm}\\}

%%%%%%%%%%%%%%%%%%%%%%%%%%%%%%%%%%%%%%%%%%%%%%%%%%

\begin{abstract}
In this paper we generalize our investigation of the unitarity of non-compact WZNW models connected to Hermitian symmetric spaces to the N=1 world-sheet supersymmetric extension of these models. We will prove that these models have a unitary spectrum in a BRST approach for antidominant highest weight representations if the level and weights of the gauged subalgebra are integers. We will find new critical string theories in 7 and 9 space-time dimensions. 
\end{abstract}

%%%%%%%%%%%%%%%%%%%%%%%%%%%%%%%%%%%%%%%%%%%%%%%%%%

\sect{Introduction}
In a previous paper \cite{Bjornsson:2007ha} we considered strings on group manifolds connected to Hermitian symmetric spaces of non-compact type. These are constructed by starting with a non-compact group $G$ which has maximal compact subgroup $H$ with a one-dimensional center $Z(H)$. Then, the coset space $G/H$ is a Hermitian symmetric space of non-compact type. All these spaces have been classified. They have $G$ being $SU(p,q)$, $SO(p,2)$, $SO^*(2n)$ and $Sp(2p,\R)$. Furthermore, there are two connected to exceptional groups, $E_{6}$ ($E_{6|-14}$) and $E_{7}$ ($E_{7|-25}$). As the maximal compact subgroup has a one-dimensional center, one can use these groups to construct non-trivial space-time backgrounds where strings can propagate. Write the maximal compact subgroup as $H = H'\times Z(H)$ and construct the coset $G/H'$. 

This background has a compact direction which plays the r{\^o}le of time. One can formulate this as a WZNW model, as was done in \cite{Hwang:1998tr} and prove it to be unitary using a BRST approach \cite{Bjornsson:2007ha}. If one assumes that the weights and level of the subalgebra are integers\footnote{The level of the algebra $\hat{\mf{g}}$ is integer, or if the Dynkin index is different from one, also half integer.}, the spectrum is unitary. This is not the case for the Goddard-Kent-Olive coset construction \cite{Bjornsson:2007ha}. In a realistic string model one would take the infinite cover so that the time is uncompactified. Furthermore, we do not consider sectors corresponding to spectral flow, which was originally proposed in \cite{Henningson:1991jc} and further elaborated in refs. \cite{Maldacena:2000hw}-\cite{Maldacena:2001km} for the $SL(2,\R)$ string. Such sectors would be crucial in constructing an S-matrix unitary string theory, as modular transformations mix different spectrally flowed sectors for the $SL(2,\R)$ string.

In this paper we will extend our analysis and consider the $\mathcal{N}=1$ world-sheet supersymmetric extension of the WZNW model \cite{Di Vecchia:1984ep} \cite{Abdalla:1984ef}. This model has bosonic and fermionic degrees of freedom. These two sectors mix in a trivial way and can be decoupled by redefining the affine generators. Constructing coset models of these theories was first done in \cite{Goddard:1984vk} for the bosonic case and a simple $\mathcal{N}=1$ supersymmetric extension of WZNW model. The generalization to other world-sheet supersymmetric WZNW models was considered in \cite{Kazama:1988qp} \cite{Kazama:1988uz} and is, therefore, called the Kazama-Suzuki coset construction. We will in this paper use a BRST approach to the coset construction \cite{Karabali:1989dk}, as this formulation gives unitarity for the bosonic case \cite{Bjornsson:2007ha}. The BRST formulation of the Kazama-Suzuki model was considered in \cite{Rhedin:1995um} and \cite{FigueroaO'Farrill:1995pv}.

The relevant representations which we consider here are so-called antidominant highest weight representations. In addition, we will assume that the level and weights of the corresponding subalgebra are integers. We will find that the $\mathcal{N}=1$ world-sheet supersymmetric extension of the WZNW-model has a unitary spectrum. At the end of the paper we will give a table of the relevant coset models. Some of these models will by themselves yield criticality. These occur in seven dimensions: $SU(3,1)/SU(3)$ at level $k=-6$ and $SO(3,2)/SO(3)$ at level $k=-6$; in nine dimensions: $SU(4,1)/SU(4)$ at level $k=-40$ and $SO(4,2)/SO(4)$ at level $k=-32$. 

In our previous paper \cite{Bjornsson:2007ha}, we also considered the necessity of our assumed representations. The proof given there is, however, not in general complete. It holds when the rank of $\mf{g}$ is two (see below for explanation of notation). We will in a forthcoming publication discuss more general cases.

Let us introduce some notations and basic definitions. We will in general follow the conventions and notations used in \cite{Fuchs:1997jv}. Denote by $\mf{g}$ and $\mf{h}$ the Lie algebras corresponding to the non-compact group $G$ and its maximal compact subgroup $H$, which admit a Hermitian symmetric space of the form $G/H$. Let $\mf{g}^{\C}$ and $\mf{h}^{\C}$ denote the corresponding complex Lie algebras. We will always take the rank $r_{\mf{g}}$ of $\mf{g}$ to be greater than one and $\mf{g}$ to be simple. One knows that $\mf{h}$ has a one-dimensional center, thus one can split $\mf{h}$ as $\mf{h}'\oplus \mf{u}(1)$. We choose $\mf{h}^{\C}$ such that it is a regular embedding in $\mf{g}^{\C}$ i.e.\ using the Cartan-Weyl basis, the Cartan elements of $\mf{h}^{\C}$, as well as generators corresponding to positive/negative roots, are all in the corresponding decomposition of $\mf{g}^{\C}$. 

Denote by $\Delta$ all roots, $\Delta^{+/-}$ the positive/negative roots, $\Delta_s$ the simple roots, $\Delta_c$ the compact roots, $\Delta_c^+=\Delta_c\cap\Delta^+$ the compact positive roots, $\Delta_n$ the non-compact roots and $\Delta_n^+$ the positive non-compact roots. We take the long roots to have length $\sqrt 2$. Let $\alpha\in \Delta$ and define the coroot by $\al^\vee=2\left(\al,\al\right)^{-1}\al$. Let $\alpha^{(i)}\in \Delta^+$ denote the simple roots. When we need to distinguish between different root systems, we denote by $\Delta^{\mf g}$ and $\Delta^{\mf h'}$ the roots in $\mf g^{\C}$ and ${\mf{h}'}^{\C}$, respectively. Furthermore, we use capital letters, $A,B,\ldots$, and small letter, $a,b,\ldots$, to denote elements in the algebra $\mf{g}$ and $\mf{h}'$ respectively. Denote by $J^A$ a generic generator of $\hat{\mf{g}}^{\C}$ which in a Cartan-Weyl basis is
\eqnb
J^i
      &=&
          H^i,\phantom{E^\al,}i=1,\ldots,r_{\mf{g}};
      \no
J^\al
      &=&
          E^\al,\phantom{H^i,}\al\in\Delta.
\eqne
We fix the basis of the root space such that the highest root is non-compact. In addition, one can choose the basis of roots such that if $\al\in\Delta^{\mf g}_c$ then the first component is zero and the other $r_{\mf{g}}-1$ components are, in general, non-zero. This, furthermore, yields an isomorphism between $\Delta_c^{\mf{g}}$ and $\Delta^{\mf{h}'}$. $g_{\mf{g}}^\vee$ denotes the dual Coxeter number of $\mf{g}$. It is well-known that one can choose a basis such that there is a unique non-compact simple root and if $\alpha\in\Delta_n^+$ then the coefficient of the non-compact simple root is always one in a simple root decomposition of $\alpha$. Dynkin diagrams and relations between positive non-compact roots of the Lie-algebras are presented in the appendix of \cite{Jakobsen:1983} and, with our notation, in \cite{Bjornsson:2007ha}. For the embedding of $\mf{h}'$ in $\mf{g}$, one defines the Dynkin index of an embedding
\eqnb
I_{\mf{h}'\subset\mf{g}}
      &=&
          \frac{\left(\theta(\mf{g}),\theta(\mf{g})\right)}{\left(\theta(\mf{h}'),\theta(\mf{h}')\right)},
\eqne
where, for regular embeddings, $\theta(\mf{h}')$ is the highest root of $\mf{h}'$ in $\mf{g}$. The conformal anomaly for the $G/H'$ WZNW model is
\eqnb
c_{\mathrm{tot}}
      &=&
          \frac{\left(k-g^{\vee}_{\mf{g}}\right)\dim\left({\mf{g}}\right)}{k}+\frac{1}{2}\dim\left({\mf{g}}\right)
          -\frac{\left(\kappa-g^{\vee}_{\mf{h}'}\right)\dim\left({\mf{h}'}\right)}{\kappa}
          -\frac{1}{2}\dim\left({\mf{h}'}\right),
\label{conformalanomaly}
\eqne
where $\kappa$ is defined as
\eqnb
\kappa&\equiv&I_{\mf{h}'\subset\mf{g}}k.
\eqne

Let us use operator product language to write down the relations for the algebra. The operator product expansions (OPE's) of the $\mathcal{N}=1$ superaffine algebra are
\eqnb
\underbracket{J^A(z)J^B(w)}
      &\sim&
          k\kappa^{AB}\left(z-w\right)^{-2}+{f^{AB}}_CJ^C(w)\left(z-w\right)^{-1}
      \\
\underbracket{J^A(z)\lambda^B(w)}
      &\sim&
          {f^{AB}}_C \lambda^C(w)\left(z-w\right)^{-1}
      \\
\underbracket{\lambda^{A}(z)\lambda^{B}(w)}
      &\sim&
          k\kappa^{AB}\left(z-w\right)^{-1}\times\left\{
          \begin{array}{l}
              1\phantom{\frac{1}{2}\left(\sqrt{\frac{z}{w}}+\sqrt{\frac{w}{z}}\right)}{\mathrm{for\; R-fermions}}\\
              \frac{1}{2}\left(\sqrt{\frac{z}{w}}+\sqrt{\frac{w}{z}}\right)\phantom{1}{\mathrm{for\; NS-fermions}}
          \end{array}
          \right.
          ,
\eqne
where $\sim$ means equality up to regular terms, ${f^{AB}}_C$ and $\kappa^{AB}$ are the structure constants and killing form, 
respectively, for the corresponding finite dimensional Lie algebra. The non-zero values in a Cartan-Weyl basis of $\mf{g}$ are 
\eqnb
{f^{i\al}}_\be
      &=&
          \al^{i}\delta_{\al,\be}
      \no
{f^{\al\be}}_i
      &=&
          \al^\vee_i\delta_{\al+\be,0}
      \no
{f^{\al\be}}_{\gamma}
      &=&
          e_{\al,\be}\delta_{\al+\be,\gamma}
      \no
\kappa^{ij}
      &=&
          G^{ij}
      \no
\kappa^{\al\be}
      &=&
          \frac{2}{\left(\al,\al\right)}\delta_{\al+\be,0},
\eqne
where $G^{ij}\equiv({\al^{(i)}}^\vee,{\al^{(j)}}^\vee)$ and $e_{\al,\be}$ is non-zero iff $\al+\be$ is a root. The correspondence between OPE's and brackets are\footnote{This is for bosonic operators and Ramond fermions.}
\eqnb
\left[A_m,B_n\right]
      &=&
          \frac{1}{\left(2\pi \ii\right)^2}\oint dw w^{\Delta_B+n-1}\oint dz z^{\Delta_A+m-1} \mathcal{R}(A(z)B(w)),
\eqne
where $\Delta$ is the conformal dimension of the operator and $\mathcal{R}$ is the radial ordering of the operators
\eqnb
\mathcal{R}(A(z)B(w))
      &=&
          \left\{
          \begin{array}{rr}
              A(z)B(w) & \left|z\right| > \left|w\right| \\
              \left(-\right)^{\epsilon_{A}\epsilon_{B}}B(w)A(z) & \left|z\right| < \left|w\right|
          \end{array}
          \right.,
\eqne
where $\epsilon$ denote the Grassman parity of the operator. The radial ordering satisfies
\eqnb
\mathcal{R}(A(z)B(w))
      &\sim&
          :A(w)B(w): + \underbracket{A(z)B(w)},
\eqne
where $:\;\;:$ means normal ordering. The element corresponding to the center, $\hat{\mf{u}}(1)$, is
\eqnb
H_m
      &=&
          {\Lambda_{(1)}}_i H^i_m,
\eqne
where ${\Lambda_{(i)}}$ are fundamental weights which are defined by $\left(\Lambda_{(i)},{\al^{(j)}}^\vee\right)=\delta_i^j$.  $H_m$ satisfies
\eqnb
\left[H_m, E^{\pm\al}_n\right]
      &=&
          0,\phantom{\pm E^{\pm\be}_{m+n},}\al\in\Delta_c^+
      \no
\left[H_m, E^{\pm\be}_n\right]
      &=&
          \pm E^{\pm\be}_{m+n},\phantom{0,}\be\in\Delta_n^+.
\eqne
One can also define new currents 
\eqnb
\hat{J}^A(z)
      &=&
          J^A(z)+\frac{1}{2k}{f^A}_{BC}:\lambda^B\lambda^C(z):
\label{eq1.14}
\eqne
which make the OPE's between $\lambda^A$ and $\hat{J}^A$ consist of only regular terms. Furthermore, these currents satisfy an affine algebra of level $k-g^\vee$.

%%%%%%%%%%%%%%%%%%%%%%%%%%%%%%%%%%%%%%%%%%%%%%%%%%

\sect{Coset construction, the BRST approach}
The Kazama-Suzuki coset construction is defined by considering states $\left|\Phi\right>$ in a representation of the supersymmetric affine algebra and then
gauging a subalgebra by the conditions
\eqnb
J^{\hat{\mf{h}}'}_+\left|\Phi\right>=\lambda^{\hat{\mf{h}}'}_+\left|\Phi\right>&=&0,
\eqne
where $J^{\hat{h}'}_+$ and $\lambda^{\hat{\mf{h}}'}_+$ are generic annihilation operators in the subalgebra $\hat{\mf{h}}'$. In the BRST approach one introduces instead a BRST operator to define the coset space \cite{Rhedin:1995um} \cite{FigueroaO'Farrill:1995pv}. One needs to introduce, in addition to the ghost sector, an auxiliary sector with currents $\tilde{J}^a(z)$ for the subalgebra $\hat{\mf{h}}'$. The BRST current is\footnote{Here we have assumed that $\mf{h}'$ is simple. If it is not one has for each simple part the BRST current as in eq.\ (\ref{BRST.coset}). Henceforth, we will present equations which hold for the case when $\mf{h}'$ is simple as the generalization to the semisimple case is straightforward.}
\eqnb
j^1_{BRST}
      &=&
          c_a\left(\hat{J}^a-\frac{1}{2k}{f^a}_{BC}\lambda^B\lambda^C+\hat{\tilde J}^a-\frac{1}{2\kappa}{f^a}_{bc}\tilde{\lambda}^b\tilde{\lambda}^c\right) + \gamma_a\left(\lambda^a+\tilde\lambda^a\right)
      \no    
      &-& 
          \frac{1}{2}{f^{ab}}_cc_ac_bb^c + {f^{ab}}_cc_a\gamma_b\beta^c,
\label{BRST.coset}
\eqne
where $c_a$ and $\gamma_a$ are fermionic and bosonic ghosts, respectively, with corresponding momenta $b^a$ and $\be^a$. They satisfy\footnote{In a mode expansion, $\gamma_{a,m}$ is chosen to have hermiticity properties $\gamma^{\dagger}_{i,m}=\gamma_{i,m}$, $\gamma^{\dagger}_{\al,m}=\gamma_{-\al,m}$,  $\be^{\dagger}_{i,m}=-\be_{i,m}$ and $\be^{\dagger}_{\al,m}=-\be_{-\al,m}$.}
\eqnb
\underbracket{c_a(z)b^b(w)}
      &\sim&
          \delta_a^b\left(z-w\right)^{-1}
      \no
\underbracket{\gamma_a(z)\be^b(w)}
      &\sim&
          \delta_a^b\left(z-w\right)^{-1}.
\eqne
The OPE of the BRST current defined in eq.\ (\ref{BRST.coset}) with itself yields only regular terms if $\kappa+\tilde \kappa = 0$. Thus, we take the level of the extra sector $\tilde{\kappa}$ to be $-\kappa$. To get the physical subspace we need to add contributions to this BRST charge arising from the superconformal symmetry. One finds 
\eqnb
j_{BRST}
      &=&
          :c_a\left(\hat{J}^a-\frac{1}{2k}{f^a}_{BC}\lambda^B\lambda^C+\hat{\tilde J}^a-\frac{1}{2\kappa}{f^a}_{bc}\tilde{\lambda}^b\tilde{\lambda}^c\right): + :\gamma_a\left(\lambda^a+\tilde\lambda^a\right): 
      \no
      &+& :cT: + :\gamma G:
          - \frac{1}{2}{f^{ab}}_c:c_ac_bb^c :
      +
          {f^{ab}}_c:c_a\gamma_b\beta^c:
       \no   
       &-& :\partial c cb:-:\gamma^2b+:\partial c \gamma\be: + \frac{1}{2}:c\gamma\partial \be: 
          - \frac{1}{2}:c\partial \gamma \be: 
      \no
      &+&
          :c\partial b^a c_a: + \frac{3}{2}:c\partial\be^a \gamma_a: 
          + \frac{1}{2}:c \be^a \partial\gamma_a: + :\partial c b^ac_a: 
      \no
      &+&
          :\partial c \be^a\gamma_a: 
      - :\gamma b^a\gamma_a: - :\gamma \partial\be^ac_a:-:\partial \gamma \be^ac_a:,
\label{BRST-current}
\eqne
where
\eqnb
T(z)
      &=&
          \frac{1}{2k}\kappa_{AB}\left(:\hat{J}^A\hat{J}^B:+ :\partial \lambda^A \lambda^B:\right)
      \no
      &+&
          \frac{1}{2\tilde \kappa}\kappa_{ab}\left(:\hat{\tilde J}^a\hat{\tilde J}^b:+ :\partial \tilde\lambda^a \tilde\lambda^b:\right)
          + T'(z)
      \\
G(z)
      &=&
          \frac{1}{k}\left(\kappa_{AB}:\hat{J}^A\lambda^B(z):-\frac{1}{6k}f_{ABC}:\lambda^A\lambda^B\lambda^C(z):\right)
      \no
      &+&
          \frac{1}{\tilde \kappa}\left(\kappa_{ab}:\hat{\tilde J}^a\tilde \lambda^b(z):
          -\frac{1}{6\tilde \kappa}f_{abc}:\tilde\lambda^a\tilde\lambda^b\tilde\lambda^c(z):\right) + G'(z),
\eqne
where $T'(z)$ and $G'(z)$ originates from a unitary superconformal algebra making the conformal anomaly equal to 15 as the corresponding BRST charge is only nilpolent for  $c=15$.

Furthermore, we have in the BRST charge introduced superconformal ghosts and corresponding momenta. These have the non-zero OPE's
\eqnb
\underbracket{c(z)b(w)}
      &\sim&
          \left(z-w\right)^{-1}
      \no
\underbracket{\gamma(z)\be(w)}
      &\sim&
          \left(z-w\right)^{-1}.
\eqne
From the BRST current one can define a few useful charges, one is the zero mode for the Virasoro algebra, 
\eqnb
L_0^{\mathrm{tot}}
      &=&
          \left[Q,b_0\right]
      \no
      &=&
          L_0 + \sum_{m\in\Z}\left(m:b_{-m}c_{m}:+\,m:b^a_{-m}c_{m,a}:\right) 
      \no
      &+& 
          \sum_{r\in\Z+\nu}\left(r:\be_{-r}\gamma_{r}:+\,r:\be^a_{-r}\gamma_{r,a}:\right) + a_g,
\eqne
and the other corresponds to the Cartan subalgebra of the Lie algebra
\eqnb
H_0^{\mathrm{tot},i}
      &=&
          \left[Q,b_0^i\right]
      \no
      &=&
          \hat{H}_0^i+\hat{\tilde H}_0^i - \sum_{m,\al\in\Delta}\frac{\left(\al,\al\right)}{4k}\al^{i}:\lambda_{-m}^{-\al}\lambda_{m}^{\al}:-\sum_{m,\al\in\Delta_c}\frac{\left(\al,\al\right)}{4\tilde\kappa}\al^{i}:\tilde{\lambda}_{-m}^{-\al}\tilde{\lambda}_{m}^{\al}:
      \no
      &+&
          \sum_{m,\al\in\Delta_c}\al^i:b^\al_{-n}c_{n,\al}:+\sum_{r\in \Z+\nu,\al\in\Delta_c}\al^i:\be^\al_{-r}\gamma_{r,\al}:,
\eqne
where $a_g=-1/2$ or $a_g=0$ is a normal ordering constant and $\nu=0$ or $\nu=1/2$ for the Neveu-Schwarz and Ramond sectors, 
respectively. 

%%%%%%%%%%%%%%%%%%%%%%%%%%%%%%%%%%%%%%%%%%%%%%%%%%

\sect{The statespace}

We consider here only discrete highest weight representations of the non-compact real forms. To be more precise, we consider the representations which have a highest weight that is antidominant. This subclass of highest weights will be defined below. The representations are constructed by using redefined currents where the world-sheet fermions and affine currents decouple. Furthermore, we work in the complex algebra and choose a specific hermitian conjugation of the operators to get representations of the relevant real form. A hermitian conjugation rule which yields the correct representations is
\eqnb
\left(\hat{E}^{\al}_{-n}\right)^{\dagger}
      &=&
          -\hat{E}^{-\al}_{n}
      \no
\left(\lambda^{\al}_{-n}\right)^{\dagger}
      &=&
          -\lambda^{-\al}_{n}
\eqne
where $\al\in\Delta_n$. The rest of the operators satisfy
\eqnb
\left(A_{-n}\right)^{\dagger}
      &=&
          A_{n}.
\eqne
The highest weight state in both the Neveu-Schwarz and Ramond sectors is of the form
\eqnb
\left|0;\mu,\tilde\mu\right>&\equiv&\left|0;\mu\right>\otimes\left|0;\tilde\mu\right>\otimes\left|0\right>_{ghost}.
\eqne
For the Neveu-Schwarz sector, we have 
\eqnb
A_m\left|0;\mu,\tilde\mu\right>
      &=&
          0,
\eqne
for $m>0$, where $A$ denotes a generic operator. For $m=0$ one has
\eqnb
A^{\al}_{0}\left|0;\mu\right> 
      &=&
          0
      \no
\hat{H}^i_{0}\left|0;\mu\right>
      &=&
          \mu^i\left|0;\mu\right>,
      \no
b^i_0\left|0\right>_{ghost}=b_{0}\left|0\right>_{ghost}
      &=&
          0
\eqne
where $\al\in \Delta_+$, $A= \left\{E,c,b\right\}$ and $i=1,\ldots,r$. We have here defined
\eqnb
c^\al_m = \frac{\left(\al,\al\right)}{2}\kappa^{\al\be}c_{m,\be}
      &=&
          c_{m,-\al}
\eqne
and similary for the bosonic ghosts. 

For the highest weight state in the Ramond sector the difference from the Neveu-Schwarz case comes from the zero modes for the world-sheet fermions and bosonic ghosts. The representations are constructed in the same way as for the bosonic generators with the exception of the fermions corresponding to the Cartan subalgebra. For these we introduce creation and annihilation operators by linear combinations
\eqnb
\lambda^{\tilde{a},\pm}_0
      &=&
          \left\{
          \begin{array}{lcl}
              \ds\frac{1}{\sqrt{2}}\left(\pm{\lambda'}^{2\tilde{a}-1}_0+\ii{\lambda'}^{2\tilde{a}}_0\right) & & \mathrm{r_{\mf{g}}}= 2\Z \\
              \ds\frac{1}{\sqrt{2}}\left(\pm{\lambda'}^{2\tilde{a}}_0+\ii{\lambda'}^{2\tilde{a}+1}_0\right) & & \mathrm{r_{\mf{g}}}= 2\Z+1
          \end{array}
          \right.
\label{Ramondsector.creationop}
\eqne
where $\tilde{a}=1,\ldots,\left[r_{\mf{g}}/2\right]$\footnote{$\left[\;\right]$ denotes the integer part.}. Here, ${\lambda'}^{i}_0$ are linear combinations of $\frac{1}{\sqrt{-k}}\lambda^{j}_0$ such that
\eqnb
\left[{\lambda'}^{i}_0,{\lambda'}^{j}_0\right]
      &=&
          -\delta^{i,j}.
\eqne
From these operators one can construct a set of states by defining a highest weight state that satisfies
\eqnb
\lambda^{\tilde{a},+}_0\left|0;\mu\right>=\lambda^{\al}_0\left|0;\mu\right>
      &=&0.
\eqne
This construction also generalizes to the bosonic ghosts,
\eqnb
\gamma^{\al}_{0}\left|0\right>_{ghost}=
\be^{\al}_{0}\left|0\right>_{ghost}=
\gamma^{\tilde{a},+}_{0}\left|0\right>_{ghost}=
\be^{\tilde{a},+}_{0}\left|0\right>_{ghost}
      &=&
          0,
\eqne
where $\gamma^{\tilde{a},+}_{0}$ and $\be^{\tilde{a},+}_{0}$ are constructed as in eq.\ (\ref{Ramondsector.creationop}), $\al\in\Delta^+_c$ and $\tilde{a}=1,\ldots,\left[r_{\mf{h}'}/2\right]$. For $r_{\mf{g}}$ odd, there is one extra component of $\lambda'^{i}$ which is left out of the above construction. This gives an extra two-fold degeneration, which is of no importance to us and will be disregarded in the following.

The highest weight state satisfies
\eqnb
H_0^{\mathrm{tot},i}\left|0;\mu,\tilde\mu\right>
      &=&
          \left(\mu^i+\tilde\mu^i+h_g^i\right)\left|0;\mu,\tilde\mu\right>
      \no
L_0^{\mathrm{tot}}\left|0;\mu,\tilde\mu\right>
      &=&
          \left(\frac{\mathcal{C}_2^{\mf{g}}}{2k}
          +\frac{\mathcal{C}_2^{\mf{h}'}}{2\tilde\kappa}
          +a_g\right)\left|0;\mu,\tilde\mu\right>,
\eqne
where $a_g$ is $-1/2$ or $0$ and $h_g^i$ is $2\rho^i_{\hat{\mf{h}}}$ or $\rho^i_\mf{g}+\rho^i_{\mf{h}'}$ for the Neveu-Schwarz and the Ramond sectors, respectively\footnote{The contribution to $h^i_g$ has different sources in the different sectors, for the Neveu-Schwarz sector it arises from the fermionic ghosts and from the Ramond sector it arises from the world-sheet fermions.}.  The states are now constructed by applying the redefined creation operators on the highest weight state. As the fermions and ghosts describe a free theory the corresponding statespaces are simple. For the redefined currents the representations are unchanged, except that the level is shifted to $k-g^\vee_{\mf{g}}$ and $\tilde{\kappa}-g^\vee_{\mf{h}}$ for the $\hat{\mf{g}}$- and $\hat{\tilde{\mf{h}}}$-sector, respectively. Thus, unitary representations with a highest weight $\hat{\tilde{\mu}}$, called dominant integral, satisfy for the 
$\hat{\tilde{\mf{h}}}'$-sector
\eqnb
\tilde{\mu}^i
      &\geq&
          0\phantom{1234}{ i=2,\ldots,r_{\mf{g}} }
      \no
\tilde\kappa 
      &\geq&
          g^\vee_{\mf{h}'}+\left(\theta_{\mf{h}'},\mu\right).
\label{Relationssectors}
\eqne
and antidominant representations for the $\hat{\mf{g}}$-sector are representations, with a highest weight $\hat{\mu}$ that satisfies
\eqnb
\mu^i
      &<&
          -\rho^i\phantom{1234}{ i=1,\ldots,r_{\mf{g}} }
      \no
k 
      &<&
          \left(\theta_{\mf{g}},\mu\right).
\eqne
The Verma modules corresponding to antidominant weights are irreducible, as can be deduced from the Shapovalov-Kac-Kazhdan determinant \cite{Sapovalov} \cite{Kac:1979fz}.

For well-known reasons, one does not consider the cohomology of the corresponding BRST charge of the current defined in eq.\ (\ref{BRST-current}), but the relative cohomology defined by the conditions
\eqnb
Q\left|\Phi\right>
      &=&
          0\phantom{,1234}\mathrm{NS \; and\; R\;sectors}
      \no
b_0^i\left|\Phi\right>
      &=&
          0\phantom{,1234}\mathrm{NS \; and\; R\;sectors}
      \no
b_0\left|\Phi\right>
      &=&
          0\phantom{,1234}\mathrm{NS \; and\; R\;sectors}
      \no
\be_0\left|\Phi\right>
      &=&
          0,\phantom{1234}{\mathrm{R\;sector\; and} \;r_{\mf{g}}\in 2\Z+1}
\label{BRST.relcohom}
\eqne
where  $i=2,\ldots,r_g$.

%%%%%%%%%%%%%%%%%%%%%%%%%%%%%%%%%%%%%%%%%%%%%%%%%%

\sect{Unitarity}
The mass shell condition is of the form
\eqnb
\left(\frac{\left(\mu,\mu+2\rho_{\mf{g}}\right)}{2k}+
\frac{\left(\tilde\mu,\tilde\mu+2\rho_{\mf{h}'}\right)}{2\tilde\kappa}\right)
+N+l'+a_g
      &=&
          0,
\label{Mass-shell}
\eqne
where $l'\geq0$ originates from some unitary SCFT. Following \cite{Bjornsson:2007ha}, define
\eqnb
\hspace{-4mm}
\chi^{\left(\hat{g},\hat{h}'\oplus SVir\right)}\left(\tau,\phi,\theta\right)
      & \equiv&
          \Tr\left[\exp\left[2\pi \ii \tau\left(L_0^{\mathrm{tot}}\right)\right]
          \exp\left[\ii\sum_{i=2}^{r_{\mf{g}}}\theta_iH^{\mathrm{tot},i}_0+\ii\phi H_0\right]
          \left(-1\right)^{\Delta N_{gh}}
          \right].
\label{Character.full}
\eqne
$\Delta N_{gh}$ is the ghostnumber of the state relative to the highest weight state for the ghosts. The trace is taken over all states in $\mathcal{H}^{\hat{g}}_{\hat\mu}\times\mathcal{H}^{\hat{\tilde{h}}'}_{\hat{\tilde{\mu}}}\times\mathcal{H}^{SCFT}_{l'} \times\mathcal{H}'^{ghost}$. $\mathcal{H}^{A}_{\ldots}$ denotes the statespace of $A$. Therefore, the character can be decomposed into separate parts
\eqnb
\chi^{\left(\hat{g},\hat{h}'\oplus SVir\right)}\left(\tau,\phi,\theta\right)
           &=&
          e^{2\pi \ii\tau a_{\mf{g}}}\chi^1\left(\tau,\phi,\theta\right)\chi^{\hat{\tilde{h}}'}\left(\tau,\theta\right),
\label{character.decom}
\eqne
where we have defined
\eqnb
\chi^1\left(\tau,\phi,\theta\right)
      &\equiv&
          \chi^{\hat{g}}\left(\tau,\phi,\theta\right)\chi^{SCFT}\left(\tau\right)\chi^{ghost}\left(\tau,\theta\right).
\eqne
As is well known \cite{Frenkel:1986dg}, the character defined in eq.\ (\ref{Character.full}) only gets contributions from non-trivial BRST invariant states. However, since we are interested in the relative cohomology, where the eigenvalues of $L_0^{\mathrm{tot}}$ and $H^{\mathrm{tot},i}_{0}$ are zero, one needs to project onto such states. We need, therefore, to consider the function
\eqnb
\int d\tau \mathcal{B}^{\left(\hat{g},\hat{h}'\oplus SVir\right)}(\tau,\phi)
      &\equiv&
          \int d\tau \int\prod_{i=2}^{r_{\mf{g}}}\left(d\theta_i\right)
          \left\{\chi^{\left(\hat{g},\hat{h}'\oplus SVir\right)}\left(\tau,\phi,\theta\right)\right\}.
\eqne
The $\tau$- and $\theta_i$-integrations are formal integrations which project onto the $\tau$- and $\theta_i$-independent part, making the character compatible with eq.\ (\ref{BRST.relcohom}). $\mathcal{B}^{\left(\hat{g},\hat{\tilde{h}}'\oplus SVir\right)}(\tau,\phi)$ is called the generalized branching function. This definition was first introduced in \cite{Hwang:1993nc} and extended in \cite{Hwang:1994yr}. As in \cite{Bjornsson:2007ha}, define another function, the signature function
\eqnb
\Sigma^{\left(\hat{\mf{g}},\,\hat{\mf{h'}}\oplus SVir\right)}&\hspace{-2mm}(\tau,\phi,\theta)&
     \no 
     &\hspace{-4mm}\equiv&
         \hspace{-8mm}
         \Tr'\left[\exp\left[2\pi\ii\tau\left(L^{\mathrm{tot}}_0\right)\right]
         \exp\left[\ii\sum_{i=2}^{r_{\mf{g}}}\theta_i H_0^{{\mathrm{tot}},i}+\ii\phi H_0 \right](-1)^{\Delta N_{gh}}\right],
\label{Signature.full}
\eqne
The prime on the trace indicates that the trace is taken with signs i.e.\ a state with positive (negative)
norm contributes with a positive (negative) sign in the trace. The corresponding coset signature function is defined as
\eqnb
\int d\tau\mathcal S^{\left(\hat{\mf{g}},\,\hat{\mf{h}}'\oplus SVir\right)}(\tau,\phi)
      &\equiv&
          \int d\tau\int\prod_{i=2}^{r_{\mf{g}}} ({d\theta_i})\hspace{1mm}\left\{ \Sigma^{\left(\hat{\mf{g}},\,
          \hat{\mf{h'}}\oplus SVir\right)}(\tau,\phi,\theta)\right\}.
\label{cosetsignature}
\eqne
The relevance of these functions to the present problem was stated is (cf.\ Lemma 2 in \cite{Bjornsson:2007ha}).
\begin{lemma}
 ${\cal H}^{Q}_{\hat{\mu}\tilde{\hat{\mu}}}$ is unitary if, and only if, 
\eqnb
\int d\tau\left[\mathcal{B}^{\left(\hat{\mf{g}},\,\hat{\mf{h}}'\oplus SVir\right)}(\tau,\phi)-
\mathcal{S}^{\left(\hat{\mf{g}},\,\hat{\mf{h}}'\oplus SVir\right)}(\tau,\phi)\right]=0.
\eqne
\label{lemma 1}
\end{lemma}
Here ${\cal H}^{Q}_{\hat{\mu}\tilde{\hat{\mu}}}$ denotes the irreducible subspace of $\mathcal{H}^{\hat{g}}_{\hat\mu}\times\mathcal{H}^{\hat{\tilde{h}}'}_{\hat{\tilde{\mu}}}\times\mathcal{H}^{SCFT}_{l'} \times\mathcal{H}'^{ghost}$ consisting of non-trivial states in the relative cohomology.

We can now state our main result.

\begin{theorem}
Let $\hat\mu$ be an antidominant weight. If $\hat{\tilde\mu}$ is a dominant integral weight then $\mathcal H^Q_{\hat{\mu}\hat{\tilde{\mu}}}$ is unitary.
\end{theorem}

Note that due to eq.\ (\ref{Relationssectors}), the components of $\hat{\mu}$ which belong to the subalgebra need to be integers since otherwise $\mathcal H^Q_{\hat{\mu}\hat{\tilde{\mu}}}$ is trivial.

\paragraph{Proof:}Let us prove this theorem using Lemma \ref{lemma 1}. We need to determine the character in each sector. As the character decomposes into separate parts, much is already known from \cite{Bjornsson:2007ha}. We need to derive the fermionic part of the $\hat{\mf{g}}$-sector and the contribution corresponding to the bosonic ghosts. Let us first consider the NS-sector. In this sector the part of the character in $\hat{\mf{g}}$ which is due to the fermions is
\eqnb
\chi^{\mf{g}\;Part.}_\mu\left(q,\theta,\phi\right)
      &=&
          \prod_{m=1}^\infty\left[\left(1+q^{m-1/2}\right)^{r_{\mf{g}}}
          \prod_{\al\in\Delta_c}\left(1+q^{m-1/2}\exp\left[\ii\left(\al,\theta\right)\right]\right)
          \right.
      \no
      &\times&
          \prod_{\al\in\Delta^+_n}\left(1+q^{m-1/2}\exp\left[\ii\phi\right]\exp\left[\ii\left(\al_\|,\theta\right)\right]\right)
      \no
      &\times&
          \left.
          \prod_{\al\in\Delta^+_n}\left(1+q^{m-1/2}\exp\left[-\ii\phi\right]\exp\left[-\ii\left(\al_\|,\theta\right)\right]\right)
          \right],
\eqne
where we have defined $q=\exp\left[2\ii \pi \tau\right]$ and $\al_\|\equiv\sum_{i=2}^{r_\mf{g}}\al^i\Lambda_{(i)}$. As the compact roots have no first component, we suppress the $\|$ notation for these roots. Together with the bosonic contribution, see \cite{Bjornsson:2007ha}, the character for the $\hat{\mf{g}}$-sector is
\eqnb
\chi^{\mf{g}}_\mu\left(q,\theta,\phi\right)
      &=&
          q^{\frac{\mathcal{C}_2^\mf{g}\left(\mu\right)}{2k}}\exp\left[\ii\left(\mu,\Theta\right)\right]
          \prod_{\al\in\Delta^+_c}\frac{1}{1-\exp\left[-\ii\left(\al,\theta\right)\right]}
      \no
      &\times&
          \prod_{\al\in\Delta^+_n}\frac{1}{1-\exp\left[-\ii\phi\right]\exp\left[-\ii\left(\al_{\|},\theta\right)\right]}
      \no
      &\times&
          \prod_{m=1}^\infty\left[\left(\frac{1+q^{m-1/2}}{1-q^m}\right)^{r_{\mf{g}}}
          \prod_{\al\in\Delta_c}\frac{1+q^{m-1/2}\exp\left[\ii\left(\al,\theta\right)\right]}{1-q^{m}\exp\left[\ii\left(\al,\theta\right)\right]}
          \right.
      \no
      &\times&
          \prod_{\al\in\Delta^+_n}\frac{1+q^{m-1/2}\exp\left[\ii\phi\right]\exp\left[\ii\left(\al_\|,\theta\right)\right]}{1-q^{m}\exp\left[\ii\phi\right]\exp\left[\ii\left(\al_\|,\theta\right)\right]}
      \no
      &\times&
          \left.
          \prod_{\al\in\Delta^+_n}\frac{1+q^{m-1/2}\exp\left[-\ii\phi\right]\exp\left[-\ii\left(\al_\|,\theta\right)\right]}{1-q^{m}\exp\left[-\ii\phi\right]\exp\left[-\ii\left(\al_\|,\theta\right)\right]}
          \right],
\label{NScharacter.gsector}
\eqne
where we have defined $(\mu,\Theta)=2(\al^{(1)},\al^{(1)})^{-1}\mu_1\phi+\sum_{i=2}^{r_\mf{g}}\mu_i\theta^i$. The signature function is determined in the same way. One finds
\eqnb
\Sigma^{\mf{g}}_\mu\left(q,\theta,\phi\right)
      &=&
          q^{\frac{\mathcal{C}_2^\mf{g}\left(\mu\right)}{2k}}\exp\left[\ii\left(\mu,\Theta\right)\right]
          \prod_{\al\in\Delta^+_c}\frac{1}{1+\exp\left[-\ii\left(\al,\theta\right)\right]}
      \no
      &\times&
          \prod_{\al\in\Delta^+_n}\frac{1}{1-\exp\left[-\ii\phi\right]\exp\left[-\ii\left(\al_{\|},\theta\right)\right]}
      \no
      &\times&
          \prod_{m=1}^\infty\left[\left(\frac{1-q^{m-1/2}}{1+q^m}\right)^{r_{\mf{g}}}
          \prod_{\al\in\Delta_c}\frac{1-q^{m-1/2}\exp\left[\ii\left(\al,\theta\right)\right]}{1+q^{m}\exp\left[\ii\left(\al,\theta\right)\right]}
          \right.
      \no
      &\times&
          \prod_{\al\in\Delta^+_n}\frac{1+q^{m-1/2}\exp\left[\ii\phi\right]\exp\left[\ii\left(\al_\|,\theta\right)\right]}{1-q^{m}\exp\left[\ii\phi\right]\exp\left[\ii\left(\al_\|,\theta\right)\right]}
      \no
      &\times&
          \left.
          \prod_{\al\in\Delta^+_n}\frac{1+q^{m-1/2}\exp\left[-\ii\phi\right]\exp\left[-\ii\left(\al_\|,\theta\right)\right]}{1-q^{m}\exp\left[-\ii\phi\right]\exp\left[-\ii\left(\al_\|,\theta\right)\right]}
          \right].
\label{NSsignature.gsector}
\eqne
For the ghost sector the difference from \cite{Bjornsson:2007ha} comes from the bosonic ghosts. Its contribution to the character is
\eqnb
\chi^{gh\;Part_1}\left(q,\theta\right)
      &=&
          \prod_{m=1}^\infty\left[
          \left(\frac{1}{1+q^{m-1/2}}\right)^{2\left(r_{\mf{g}}-1\right)}
          \prod_{\al\in\Delta_c}\left(\frac{1}{1+q^{m-1/2}\exp\left[\ii\left(\al,\theta\right)\right]}\right)^2
          \right].
      \no
\eqne
The bosonic superconformal ghosts give
\eqnb
\chi^{gh\;Part_2}\left(q\right)
      &=&
          \prod_{m=1}^\infty\left(\frac{1}{1+q^{m-1/2}}\right)^{2}.
\eqne
Putting things together, one ends up with the following expression for the ghost sector of the character 
\eqnb
\chi^{gh}\left(q,\theta\right)
      &=&
          \exp\left[2\ii\left(\rho_{\mf{h}'},\theta\right)\right]
          \prod_{\al\in\Delta_c^+}\left(1-\exp\left[-\ii\left(\al,\theta\right)\right]\right)^2
      \no
      &\times&
          \prod_{m=1}^\infty\left[
          \left(\frac{1-q^m}{1+q^{m-1/2}}\right)^{2r_{\mf{g}}}
          \prod_{\al\in\Delta_c}\left(\frac{1-q^m\exp\left[\ii\left(\al,\theta\right)\right]}{1+q^{m-1/2}\exp\left[\ii\left(\al,\theta\right)\right]}\right)^2
          \right]
\label{NScharacter.ghostsector}
\eqne
The signature function is determined in the same way and the final result for the ghost sector is
\eqnb
\Sigma^{gh}\left(q,\theta,\phi\right)
      &=&
          \exp\left[2\ii\left(\rho_{\mf{h}'},\theta\right)\right]
          \prod_{\al\in\Delta_c^+}\left(1-\exp\left[-\ii\left(\al,\theta\right)\right]\right)
          \left(1+\exp\left[-\ii\left(\al,\theta\right)\right]\right)
      \no
      &\times&
          \prod_{m=1}^\infty\left[
          \left(\frac{1-q^m}{1+q^{m-1/2}}\right)^{r_{\mf{g}}}\left(\frac{1+q^m}{1-q^{m-1/2}}\right)^{r_{\mf{g}}}
          \right.
      \no
      &\times&
          \left.
          \prod_{\al\in\Delta_c}
          \left(\frac{1-q^m\exp\left[\ii\left(\al,\theta\right)\right]}{1+q^{m-1/2}\exp\left[\ii\left(\al,\theta\right)\right]}\right)
          \left(\frac{1+q^m\exp\left[\ii\left(\al,\theta\right)\right]}{1-q^{m-1/2}\exp\left[\ii\left(\al,\theta\right)\right]}\right)
          \right].
\label{NSsignature.ghostsector}
\eqne
Putting eqs.\ (\ref{NScharacter.gsector}) and (\ref{NScharacter.ghostsector}) together we get
\eqnb
\chi^1_\mu\left(q,\phi,\theta\right)
      &=&
          q^{\frac{\mathcal{C}_2^\mf{g}\left(\mu\right)}{2k}}\exp\left[\ii\left(\mu,\Theta\right)+2\ii\left(\rho_{\mf{h}'},\theta\right)\right]
      \no
      &\times&
          \prod_{\al\in\Delta_c^+}
          \left(1-\exp\left[-\ii\left(\al,\theta\right)\right]\right)
          \prod_{\al\in\Delta^+_n}
          \frac{1}{1-\exp\left[-\ii\phi\right]\exp\left[-\ii\left(\al_{\|},\theta\right)\right]}
      \no
      &\times&
          \prod_{m=1}^\infty
          \left[
          \left(\frac{1-q^m}{1+q^{m-1/2}}\right)^{r_{\mf{g}}}
          \prod_{\al\in\Delta_c}
          \left(\frac{1-q^m\exp\left[\ii\left(\al,\theta\right)\right]}{1+q^{m-1/2}\exp\left[\ii\left(\al,\theta\right)\right]}\right)
          \right.
      \no
      &\times&
          \prod_{\al\in\Delta^+_n}
          \frac{1+q^{m-1/2}\exp\left[\ii\phi\right]\exp\left[\ii\left(\al_\|,\theta\right)\right]}{1-q^{m}\exp\left[\ii\phi\right]\exp\left[\ii\left(\al_\|,\theta\right)\right]}
      \no
      &\times&
          \left.
          \prod_{\al\in\Delta^+_n}
          \frac{1+q^{m-1/2}\exp\left[-\ii\phi\right]\exp\left[-\ii\left(\al_\|,\theta\right)\right]}{1-q^{m}\exp\left[-\ii\phi\right]\exp\left[-\ii\left(\al_\|,\theta\right)\right]}.
          \right]
\label{NSsector.character1}
\eqne
Doing the same for the signature function, i.e. putting eqs.\ (\ref{NSsignature.gsector}) and (\ref{NSsignature.ghostsector}) together, yields the same expression as in eq.\ (\ref{NSsector.character1}). Thus, due to Lemma \ref{lemma 1} we get unitarity.

Let us now determine the character in the Ramond sector. This problem is a bit harder due to the degeneracy of the vacuum. The states in the relative cohomology are defined in eq.\ (\ref{BRST.relcohom}). There is a degeneracy caused by $\gamma_0$ (if $r_\mf{g}$ is even), $\gamma^{\tilde{a},-}_0$ and $\be^{\tilde{a},-}_0$. If one computes the character and the signature function copying the Neveu-Schwarz case, one will find that they are ill-defined and in the form of an alternating sum or zero times infinity. Therefore, one has to introduce some regulator. Let us show how one can do this in a simpler problem, namely the world-sheet supersymmetric flat case. Let us make a simplification and treat only the reducible Majorana representation of $\mf{so}(9,1)$. Introduce a charge $N'\equiv\sum_{a}\lambda_0^{a,-}\lambda_0^{a,+}-\gamma_0\be_0$ which counts the number of excitations of $\lambda_0^{a,-}$ and $\gamma_0$. Let us introduce this charge into the trace in the character and signature functions as
\eqnb
{\Tr}\left[\exp\left[\ii\varphi N'\right]\ldots\right],
\eqne
where $\ldots$ indicates BRST-invariant operators\footnote{In this case $L_0$ and momentum operators.}. Computing the character corresponding to the zero mode of the fermions yields
\eqnb
\left(1+\exp\left[\ii\varphi\right]\right)^{5}
\eqne
and the part corresponding to the zero mode of the bosonic ghost is
\eqnb
\frac{1}{1+\exp\left[\ii\varphi\right]}.
\eqne
Observe that this result is not BRST-invariant because the charge $N'$ is not BRST-invariant.  But we can recover the BRST-invariant piece by taking the limit $\varphi\rightarrow 0$. Thus, the insertion of $\exp\left[\ii\varphi N'\right]$ in the trace acts as a regulator of the character. In the end, the character is well-defined and the result is $2^4$, which is the dimension of the sum of the two irreducible spinor representations of $\mf{so}(8)$ that the Majorana representation consists of. One can do the same for the signature function to get
\eqnb
\underbrace{\left(1+\exp\left[\ii\varphi\right]\right)^{4}\left(1-\exp\left[\ii\varphi\right]\right)}_{\mathrm{fermionic\;part}}
\underbrace{\frac{1}{1-\exp\left[\ii\varphi\right]}}_{\mathrm{bosonic\;ghost\;part}},
\eqne
which yields the same expansion as for the character. 

Let us now adapt this to the present case. Introduce the charge
\eqnb
N
      &=&
          \sum_{a=1}^{[r_{\mf{g}}/2]}\lambda^{a,-}_0\lambda^{a,+}_0
          +\sum_{a=1}^{[r_{\mf{h}'}/2]}\left(\tilde\lambda^{a,-}_0\tilde\lambda^{a,+}_0
          -\gamma^{a,-}_0\be^{a,+}_0+\be^{a,-}_0\gamma^{a,+}_0\right)-\gamma_0\be_0.
\eqne
This charge will be introduced into the trace in the character and signature functions as $\exp\left[\ii\varphi N\right]$. The character for the $\hat{\mf{g}}$ part due to the fermions is
\eqnb
\chi^{\mf{g}\;Part}_\mu\left(q,\phi,\theta,\varphi\right)
      &=&
          \exp\left[\ii\left(\rho_{\mf{g}},\Theta\right)\right]\left(1+\exp\left[\ii\varphi\right]\right)^{\left[r_{\mf{g}}/2\right]}
      \no
      &\times&
          \prod_{\al\in\Delta_c^+}\left(1+\exp\left[-\ii\left(\al,\theta\right)\right]\right)
          \prod_{\al\in\Delta^+_n}
          \left(1+\exp\left[-\ii\phi\right]\exp\left[-\ii\left(\al_{\|},\theta\right)\right]\right)
      \no
      &\times&
          \prod_{m=1}^\infty
          \left[
          \left(1+q^m\right)^{r_{\mf{g}}}
          \prod_{\al\in\Delta_c}
          \left(1+q^m\exp\left[\ii\left(\al,\theta\right)\right]\right)
          \right.
      \no
      &\times&
          \prod_{\al\in\Delta^+_n}
          \left(1+q^{m}\exp\left[\ii\phi\right]\exp\left[\ii\left(\al_\|,\theta\right)\right]\right)
      \no
      &\times&
          \left.
          \prod_{\al\in\Delta^+_n}
          \left(1+q^{m}\exp\left[-\ii\phi\right]\exp\left[-\ii\left(\al_\|,\theta\right)\right]\right)
          \right]
\eqne
Putting this together with the known bosonic contribution yields
\eqnb
\chi^{\mf{g}}_\mu\left(q,\phi,\theta,\varphi\right)
      &=&
          q^{\frac{\mathcal{C}_2^\mf{g}\left(\mu\right)}{2k}}\exp\left[\ii\left(\mu+\rho_{\mf{g}},\Theta\right)\right]
          \left(1+\exp\left[\ii\varphi\right]\right)^{\left[r_{\mf{g}}/2\right]}
      \no
      &\times&
          \prod_{\al\in\Delta_c^+}
          \frac{1+\exp\left[-\ii\left(\al,\theta\right)\right]
          }{1-\exp\left[-\ii\left(\al,\theta\right)\right]}
          \prod_{\al\in\Delta^+_n}
          \frac{1+\exp\left[-\ii\phi\right]\exp\left[-\ii\left(\al_{\|},\theta\right)\right]
          }{1-\exp\left[-\ii\phi\right]\exp\left[-\ii\left(\al_{\|},\theta\right)\right]}
      \no
      &\times&
          \prod_{m=1}^\infty
          \left[
          \left(\frac{1+q^m}{1-q^m}\right)^{r_{\mf{g}}}
          \prod_{\al\in\Delta_c}
          \frac{1+q^m\exp\left[\ii\left(\al,\theta\right)\right]}{1-q^m\exp\left[\ii\left(\al,\theta\right)\right]}
          \right.
      \no
      &\times&
          \prod_{\al\in\Delta^+_n}
          \frac{1+q^{m}\exp\left[\ii\phi\right]\exp\left[\ii\left(\al_\|,\theta\right)\right]}{1-q^{m}\exp\left[\ii\phi\right]\exp\left[\ii\left(\al_\|,\theta\right)\right]}
      \no
      &\times&
          \left.
          \prod_{\al\in\Delta^+_n}
          \frac{1+q^{m}\exp\left[-\ii\phi\right]\exp\left[-\ii\left(\al_\|,\theta\right)\right]}{1-q^{m}\exp\left[-\ii\phi\right]\exp\left[-\ii\left(\al_\|,\theta\right)\right]}
          \right].
\label{Rcharacter.gsector}
\eqne
The character corresponding to the bosonic coset ghosts is
\eqnb
\chi^{gh\;Part_1}_\mu\left(q,\theta,\varphi\right)
      &=&
          \frac{1}{\left(1+\exp\left[\ii\varphi\right]\right)^{2\left[r_{\mf{h}'}/2\right]}}
      \no
      &\times&
          \prod_{\al\in\Delta_c^+}
          \frac{1}{\left(1+\exp\left[-\ii\left(\al,\theta\right)\right]\right)^2}
      \no
      &\times&
          \prod_{m=1}^\infty
          \left[
          \frac{1}{\left(1+q^m\right)^{2\left(r_{\mf{g}}-1\right)}}
          \prod_{\al\in\Delta_c}
          \frac{1}{\left(1+q^m\exp\left[\ii\left(\al,\theta\right)\right]\right)^2}
          \right],
\eqne
and to the superconformal bosonic ghosts
\eqnb
\chi^{gh\;Part_2}_\mu\left(q,\varphi\right)
      &=&
          \frac{1}{\left(1+\exp\left[\ii\varphi\right]\right)^{\left[r_{\mf{g}}/2\right]-\left[r_{\mf{h}'}/2\right]}}
          \prod_{m=1}^\infty\frac{1}{\left(1+q^m\right)^{2}},
\eqne
which together with the known contributions of the fermionic ghosts, yields the character for the full ghost sector
\eqnb
\chi^{gh}_\mu\left(q,\theta,\varphi\right)
      &=&
          \frac{1}{\left(1+\exp\left[\ii\varphi\right]\right)^{\left[r_{\mf{g}}/2\right]+\left[r_{\mf{h}'}/2\right]}}
          \prod_{\al\in\Delta_c^+}
          \left(\frac{1-\exp\left[-\ii\left(\al,\theta\right)\right]}{1+\exp\left[-\ii\left(\al,\theta\right)\right]}
          \right)^2
      \no
      &\times&
          \prod_{m=1}^\infty
          \left[
          \left(\frac{1-q^m}{1+q^m}\right)^{2r_{\mf{g}}}
          \prod_{\al\in\Delta_c}
          \left(\frac{1-q^m\exp\left[\ii\left(\al,\theta\right)\right]}{1+q^m\exp\left[\ii\left(\al,\theta\right)\right]}\right)^2
          \right].
\label{Rcharacter.ghostsector}
\eqne
Putting together eqs.\ (\ref{Rcharacter.gsector}) and (\ref{Rcharacter.ghostsector}) one will get the character for the 
$\hat{\mf{g}}$- and ghost sectors
\eqnb
\chi^1_\mu\left(q,\phi,\theta,\varphi\right)
      &=&
          q^{\frac{\mathcal{C}_2^\mf{g}\left(\mu\right)}{2k}}\exp\left[\ii\left(\mu+\rho_{\mf{g}},\Theta\right)\right]
      \no
      &\times&
          \frac{1}{\left(1+\exp\left[\ii\varphi\right]\right)^{\left[r_{\mf{h}'}/2\right]}}
          \prod_{\al\in\Delta_c^+}
          \frac{1-\exp\left[-\ii\left(\al,\theta\right)\right]}{1+\exp\left[-\ii\left(\al,\theta\right)\right]}
      \no
      &\times&
          \prod_{\al\in\Delta^+_n}
          \frac{1+\exp\left[-\ii\phi\right]\exp\left[-\ii\left(\al_{\|},\theta\right)\right]}
          {1-\exp\left[-\ii\phi\right]\exp\left[-\ii\left(\al_{\|},\theta\right)\right]}
      \no
      &\times&
          \prod_{m=1}^\infty
          \left[
          \left(\frac{1-q^m}{1+q^{m}}\right)^{r_{\mf{g}}}
          \prod_{\al\in\Delta_c}
          \left(\frac{1-q^m\exp\left[\ii\left(\al,\theta\right)\right]}{1+q^{m}\exp\left[\ii\left(\al,\theta\right)\right]}\right)
          \right.
      \no
      &\times&
          \prod_{\al\in\Delta^+_n}
          \frac{1+q^{m}\exp\left[\ii\phi\right]\exp\left[\ii\left(\al_\|,\theta\right)\right]}{1-q^{m}\exp\left[\ii\phi\right]\exp\left[\ii\left(\al_\|,\theta\right)\right]}
      \no
      &\times&
          \left.
          \prod_{\al\in\Delta^+_n}
          \frac{1+q^{m}\exp\left[-\ii\phi\right]\exp\left[-\ii\left(\al_\|,\theta\right)\right]}{1-q^{m}\exp\left[-\ii\phi\right]\exp\left[-\ii\left(\al_\|,\theta\right)\right]}
          \right].
\label{Rcharacter1}
\eqne

Let us now determine the corresponding signature function. This is done in the same way as for the character. By a straightforward calculation one finds
\eqnb
\Sigma^{\mf{g}}_\mu\left(q,\phi,\theta,\varphi\right)
      &=&
          q^{\frac{\mathcal{C}_2^\mf{g}\left(\mu\right)}{2k}}\exp\left[\ii\left(\mu+\rho_{\mf{g}},\Theta\right)\right]
          \left(1-\exp\left[\ii\varphi\right]\right)^{\left[r_{\mf{g}}/2\right]}
      \no
      &\times&
          \prod_{\al\in\Delta_c^+}
          \frac{1-\exp\left[-\ii\left(\al,\theta\right)\right]
          }{1+\exp\left[-\ii\left(\al,\theta\right)\right]}
          \prod_{\al\in\Delta^+_n}
          \frac{1+\exp\left[-\ii\phi\right]\exp\left[-\ii\left(\al_{\|},\theta\right)\right]
          }{1-\exp\left[-\ii\phi\right]\exp\left[-\ii\left(\al_{\|},\theta\right)\right]}
      \no
      &\times&
          \prod_{m=1}^\infty
          \left[
          \left(\frac{1-q^m}{1+q^m}\right)^{r_{\mf{g}}}
          \prod_{\al\in\Delta_c}
          \frac{1-q^m\exp\left[\ii\left(\al,\theta\right)\right]}{1+q^m\exp\left[\ii\left(\al,\theta\right)\right]}
          \right.
      \no
      &\times&
          \prod_{\al\in\Delta^+_n}
          \frac{1+q^{m}\exp\left[\ii\phi\right]\exp\left[\ii\left(\al_\|,\theta\right)\right]}{1-q^{m}\exp\left[\ii\phi\right]\exp\left[\ii\left(\al_\|,\theta\right)\right]}
      \no
      &\times&
          \left.
          \prod_{\al\in\Delta^+_n}
          \frac{1+q^{m}\exp\left[-\ii\phi\right]\exp\left[-\ii\left(\al_\|,\theta\right)\right]}{1-q^{m}\exp\left[-\ii\phi\right]\exp\left[-\ii\left(\al_\|,\theta\right)\right]}
          \right].
\label{Rsignature.gsector}
\eqne
One can proceed in the same way for the ghost sector and end up with the expression 
\eqnb
\Sigma^{gh}_\mu\left(q,\theta, \varphi\right)
      &=&
          \frac{1}{\left(1+\exp\left[\ii\varphi\right]\right)^{\left[r_{\mf{h}'}/2\right]}
          \left(1-\exp\left[\ii\varphi\right]\right)^{\left[r_{\mf{g}}/2\right]}}.
\label{Rsignature.ghostsector}
\eqne

If one puts eqs.\ (\ref{Rsignature.gsector}) and (\ref{Rsignature.ghostsector}) together determining $\Sigma^1_\mu\left(q,\phi,\theta, \varphi\right)$ one discovers that it is equal to eq.\ (\ref{Rcharacter1}), even before the limit $\varphi\rightarrow 0$. This implies that if the signature function and the character of the $\tilde{\hat{h}}'$ sector are equal, then the full character and signature function are equal. By Lemma \ref{lemma 1} this gives unitarity. As this occurs when the representation of $\tilde{\hat{h}}'$ is unitary, i.e.\ when $\hat{\tilde{\mu}}$ is dominant integral, we have now proven the theorem. $\Box$

Let now consider the converse of the theorem, i.e.\ that it is necessary that the representation of the auxiliary sector is dominant integral to have unitarity. As remarked in the introduction, the proof presented in \cite{Bjornsson:2007ha} is not complete for higher rank theories. For $r_{\mf{g}}=2$ i.e. the cases $SU(2,1)/SU(2)$ and $SL(4,\R)/SU(2)$, however, the proof holds. This gives immediate implications to the present case, as the bosonic sector is part of the supersymmetric model. Thus, we get the following.
\begin{theorem}
Let $\left(\mf{g},\mf{h}'\right)$ be either $\left(\mf{su}(2,1),\mf{su}(2)\right)$ or $\left(\mf{sl}(4,\R),\mf{su}(2)\right)$, $\hat{\mu}$ be an antidominant weight and assume $\mathcal H^Q_{\hat{\mu}\hat{\tilde{\mu}}}$ to be non-trivial. If $\hat{\tilde{\mu}}$ is not a dominant weight then there will always exist a state in $\mathcal H^Q_{\hat{\mu}\hat{\tilde{\mu}}}$ which is non-unitary.
\end{theorem}

It is straightforward, using similar expressions as the ones above for the characters and signature functions, to prove unitarity for the supersymmetric coset $G/H$ i.e.\  when we gauge the full compact subalgebra but not imposing the super Virasoro constraints. This is a generalization of the results for the bosonic case proven in \cite{Bjornsson:2007ha}. The fact that we get unitarity for these cases is natural to expect, as we no longer have any time-like excitations. We will present a more detailed report on this and further similar cases in a forthcoming publication.

\begin{table}[h!]
\caption{}
\label{Table:CFT}
\vspace{0.3cm}

\begin{tabular}{ccccccccc}
Coset $\mf{g};\;\mf{h}'$ & $\dim\left(\mf{g}\right)$ & $\dim\left(\mf{h}'\right)$ & $g^\vee_{\mf{g}}$ & $g^\vee_{\mf{h}'}$ & $c_{\max}$ &  $k_{\max}$ & $c_{\min}$ & $d$\\
\hline
$\mf{su}(2,1);\;{\mf{su}(2)}$                 & 8  & 3          & 3 & 2 & 13,5    & -3       & 7.5   & 5   \\
$\mf{su}(3,1);\;{\mf{su}(3)}$                 & 15 & 8          & 4 & 3 & 15      & -6       & 10.5  & 7   \\
$\mf{su}(4,1);\;{\mf{su}(4)}$                 & 24 & 15         & 5 & 4 & 15      & -40      & 13.5  & 9   \\
$\mf{su}(2,2);\;{\mf{su}(2)\oplus\mf{su}(2)}$ & 15 & 2$\times$3 & 4 & 2 & 15      & -32      & 13.5  & 9   \\
$\mf{so}(3,2);\;\mf{so}(3)$                   & 10 & 3          & 3 & 2 & 15      & -6       & 10.5  & 7   \\
$\mf{so}(4,2);\;\mf{so}(4)$                   & 15 & 6          & 4 & 2 & 15      & -32      & 13.5  & 9   \\
$\mf{so}^*(6);\;\mf{su}(3)$                   & 15 & 8          & 4 & 3 & 15      & -6       & 10.5  & 7   \\
$\mf{sp}(4,\R);\;\mf{su}(2)$                  & 10 & 3          & 3 & 2 & 15      & -6       & 10.5  & 7   \\
\end{tabular}
\end{table}

In Table \ref{Table:CFT} we present the different world-sheet supersymmetric string theories which have a unitary spectrum. As the level of the subalgebra is an integer one can have a largest possible value of the conformal anomaly which satisfies $c\leq 15$, as can be seen from eq.\ (\ref{conformalanomaly}). This value of $c$ is called $c_{\max}$ and the corresponding value of the level is called $k_{\max}$ in table \ref{Table:CFT}. In the table we also give the minimal value of the conformal anomaly which arises in the limit $k\rightarrow -\infty$. Furthermore, we give the number of dimensions of the space-time background, denoted by $d$. Note that one always gets an odd dimension as the space-like degrees of freedom are associated with the noncompact roots, which come in pairs.

In this table there are a few isomorphisms, $\mf{so}(4,2)\cong \mf{su}(2,2)$, $\mf{so}^*(6)\cong\mf{su}(3,1)$, $\mf{so}(3,2)\cong\mf{sp}(4,\R)$, $\mf{so}(3)\cong\mf{su}(2)$ and $\mf{so}(4)\cong\mf{su}(2)\oplus\mf{su}(2)$.

%%%%%%%%%%%%%%%%%%%%%%%%%%%%%%%%%%%%%%%%%%%%%%%%%%%%%%%%%%
\vspace{0.5cm}

\noindent
{\bf{Acknowledgements}}
\\
\noindent
S.H. is partially supported by the Swedish Research Council under project no.\ 621-2005-3424.

\end{document}